\documentclass{an}
\usepackage{graphicx}
\usepackage{times}
\usepackage{fancyhdr}
\begin{document}

\title{Variability in the  Lambda Orionis cluster substellar domain\thanks{Observations
collected at the WHT, Roque de los Muchachos Observatory, Spain}
}

\author{D. Barrado y Navascu\'es\inst{1} 
\and  N. Hu\'elamo\inst{2}
\and M. Morales Calder\'on\inst{1}}
\institute{
Laboratorio de Astrof\'{\i}sica Espacial y F\'{\i}sica Fundamental (LAEFF-INTA), 
Apdo. 50727, 28080 Madrid, Spain
\and 
ESO, Alonso de Cordova 3107, Santiago, Chile
}

\date{Received $<$date$>$; 
accepted $<$date$>$;
published online $<$date$>$}

\abstract{
We present the first results on variability of very low mass stars and brown dwarfs
belonging to the $\sim$5 Myr Lambda Orionis cluster (Collinder 69). We have 
monitored almost continuously in the $J$ filter a small 
area of the cluster which includes 12 possible members
of the cluster during one night. Some members have turned to be short-term
variable. One of them, LOri167, has a mass close to the planetary mass limit and 
its variability might be due to instabilities produced by the deuterium burning,
although other mechanism cannot be ruled out.
}
\correspondence{barrado@laeff.esa.es}

\maketitle


\section{Introduction}
We are embarked in a very ambitious project devoted to the Lambda
Orionis Star Forming Region (SFR), the Head of the Orion giant, 
the Hellenic mythological hero.  This
area contains several distinct structures, including several dark
clouds with active star formation and a very young cluster (Collinder
69 or Lambda Orionis cluster) associated to the multiple star
$\lambda$ Ori, located at the center of the SFR.  Additional details
can be found in Dolan \& Mathieu (1999, 2001) and Barrado y
Navascu\'es et al. (2004, 2005). This paper
deals with a specific topic: variability at the bottom of the stellar
sequence in the Lambda Orionis cluster. Specifically, we are interested on the
faintest object in our sample, trying to establish whether the
variability arises from pulsation due to deuterium burning, a
mechanism initially proposed Toma (1972) --based on previous work by by Gabriel (1964)--
 for pre-main sequence stars (2.0-0.2 M$_\odot$) and recently suggested by Palla \& Baraffe
(2005) for young brown dwarfs.

So far, few young brown dwarfs have been photometrically monitored,
and their periods derived (see Scholz et al. 2005, this volume).
The pioneering work dealt with a handful of very low mass members 
of the Pleiades cluster (Terndrup et al. 1999; see also Scholz \& Eisl\"offel 2004a).
 The youngest objects belong to Chamaeleon I (1--3Myr),
 and they have the longest periods, in the range of 53-114 h and
with amplitudes in the $J$ band of 0.13-0.05 mag (Joergens et al.
2003). Several works have focused on brown dwars of the somewhat older (3--5
Myr) $\sigma$ Orionis cluster (Bailer-Jones \& Mundt 2001; Caballero et
al.  2004; Scholz \& Eisl\"offel 2004b). They have unveiled periods in
the range 3-240 h, with amplitudes of 0.02-0.4   mag. However, the
lowest mass object whose variability has been detected  so far, S\,Ori\,45,
 has a  mass  about 0.02 M$_\odot$. Its period   has a been estimated as 
0.50$\pm$0.12 h (Bailer-Jones \& Mundt 2001). 
Later on, Zapatero-Osorio et al. (2003) derived two possible periods,
either $\sim$0.77 h or a value in the range   2.5--3.6h,
 with amplitudes in the range 0.01-0.13 mag.
Because of escape velocity arguments, the shortest period cannot be
the rotational period. Palla \& Baraffe (2005) suggested the
possibility of pulsations due to deuterium burning. However, except
for S\,Ori\,45, and taken into account the age, the other objects seem to
be too massive (larger than 0.05 M$_\odot$) to be undergoing deuterium
burning (ie., they should lack pulsations driven by its combustion).
We will discuss the observations and data analysis in the next sections.

\section{Analysis}

\subsection{The observations}

The data presented in this paper were collected during two campaigns
in November 19-20 2002 and February 12-14 2003 at the 4.2m WHT, La
Palma (Spain), and the near infrared camera INGRID, which has a
4.2$\times$4.2 arcmin$^2$ field of view.  During the first run, we divided the area
around the Lambda Ori cluster in a 42$\times$28 arcmin$^2$ grid, composed
of squares of 4$\times$4 arcmin each, trying to match our CFHT optical
survey (Barrado y Navascu\'es et al. 2004). We used the $J$ filter, but
in few cases we took additional data with the filters $H$ and $Ks$.  The
exposure times were five individual exposures (with a small offset)
for each pointing of 60 seconds each.  Therefore, the total exposure
time was five minutes per pointing.  The same procedure was repeated
during the 2003 run. These observations will be reported elsewhere.
The data were processed using IRAF\footnote{IRAF is distributed by
National Optical Astronomy Observatories, which is operated by the
Association of Universities for Research in Astronomy, Inc., under
contract to the National Science Foundation, USA}.

\setcounter{figure}{0}
    \begin{figure}
    \centering
    \includegraphics[width=8.2cm]{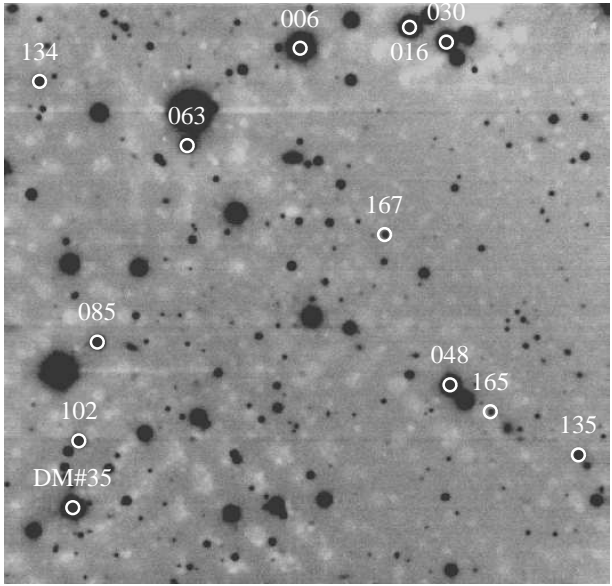}
 \caption{
Deep INGRID image in the $J$ band of the area southeast of the star $\lambda$ Orionis.
We have marked the cluster candidates from Dolan \& Mathieu (1999) and 
Barrado y Navascu\'es et al. (2004).
}
 \end{figure}

\setcounter{table}{0}
\begin{table*} 
\caption{Photomotry of our candidate members.}
\begin{tabular}{lccccccccc}
 Name            & $Rc$ error & $Ic$ error& $J$$^1$error &$J$$^2$error&$H$$^2$error &$Ks$$^2$error&Mass$^6$ (M$_\odot$)&Sp.T.& Var?\\
\hline
[DM99] 35$^3$   & 14.48 0.01 & 13.61 0.01 & 12.136 0.001 & 12.26 0.01 & 11.60  0.01 & 11.55 0.01  & 0.785 0.739 0.571 &     & \\  
LOri006         & 13.55 0.01 & 12.75 0.01 & 11.430 0.001 & 11.67 0.01 & 10.90  0.01 & 10.94 0.01  & 1.106 1.171 0.901 &     & \\   
LOri016         & 14.07 0.01 & 13.18 0.01 & 11.902 0.001 & 12.00 0.01 & 11.41  0.01 & 11.27 0.01  & 0.970 0.875 0.696 &     & \\  
LOri030         & 14.95 0.01 & 13.74 0.01 & 12.314 0.001 & 12.44 0.01 & 11.81  0.01 & 11.64 0.01  & 0.735 0.655 0.537 &     & \\  
LOri048         & 15.78 0.01 & 14.41 0.01 & 12.622 0.002 & 12.78 0.01 & 12.17  0.01 & 12.00 0.01  & 0.498 0.534 0.418 &     & \\  
LOri063$^4$     & 16.80 0.01 & 15.34 0.01 & 13.610 0.002 & 13.72 0.01 & 13.02  0.01 & 12.69 0.01  & 0.287 0.269 0.261 & M4.5& Y \\   
LOri085$^4$     & 17.65 0.01 & 16.04 0.01 & 14.058 0.003 & 14.21 0.01 & 13.58  0.01 & 13.26 0.01  & 0.184 0.194 0.170 &     & N \\  
LOri102         & 18.24 0.01 & 16.50 0.01 & 14.430 0.004 & 14.57 0.01 & 14.05  0.01 & 13.78 0.01  & 0.137 0.147 0.113 &     & \\   
LOri134         & 19.91 0.01 & 17.90 0.01 & 15.609 0.006 & 15.72 0.01 & 15.17  0.01 & 14.82 0.01  & 0.063 0.059 0.046 & M5  & Y \\  
LOri135         & 19.91 0.01 & 17.90 0.01 & 15.529 0.006 & 15.63 0.01 & 15.14  0.01 & 14.79 0.01  & 0.063 0.063 0.048 & M7  & Y \\  
LOri165$^5$     & 23.12 0.22 & 20.73 0.02 & 18.497 0.026 & 18.57 0.05 & 18.08  0.05 & 17.83 0.09  &       --          &     & \\  
LOri167         & 23.86 0.64 & 20.90 0.02 & 17.776 0.017 & 17.84 0.03 & 17.15  0.03 & 16.62 0.03  & 0.018 0.014 0.015 &     & Y \\  
\hline
\end{tabular}
$\,$\\
$^1$ INGRID  Feb. 12, 2003. \\
$^2$ INGRID  Nov. 19, 2002. \\
$^3$ From Dolan \& Mathieu (1999), the spectral type is in the range K8--M2, from the colors.\\
$^4$ Class II object, based on Spitzer/IRAC data (Barrado y Navascu\'es et al. 2006).\\
$^5$ Possible non-member.\\
$^6$ As derived from $I$, $J$ and $K$ magnitudes, $(m-M)_0$=8.010, $E(B-V)$=0.12 mag and 5 Myr isochrones from the Lyon group.\\
\end{table*}

Here, we analyze the data taken just southeast of the $\lambda$
Orionis star.  In 2002 we collected $JHKs$ imaging of this pointing,
whereas in 2003 we devoted a whole night, with a five-hour time-span,
in order to observe this small area in the $J$ band, looking for short term
variability.  The individual 60 seconds exposures were processed,
aligned and combined into one single 5 minutes frame.  In total, we
collected 51 of such five-minute images.  In addition, we combined
all the frames into a single image in order to derive deep, accurate
photometry.  The final deep image can be seen in Figure 1, where we
have marked the location of our targets.  The star BD $+$09 879C, the
third component of the $\lambda$ Ori system (30 arcsec south of the
primary), can be partially  seen at the upper-right corner of the finding chart.
Optical and near infrared photometry, as well as a mass estimate based
on a 5 Myr isochrones by Baraffe et al. (1998) are listed in Table 1.


Note that we have previously collected 
 low resolution spectroscopy for two of the objects
listed in Table 1, LOri 134 and 135. The spectral types are M5 and M7,
and they were classified as probable non-member and possible member,
respectively, based on their spectral type and the optical and 2MASS
photometry (less accurate than our new values). Based on the new
photometry (including Spitzer/IRAC data, see Barrado y Navascu\'es et
al. 2006, in preparation), we have classified them as cluster
members. The other objects analyzed here, except for LOri165, seem to
be members of the association too.

\setcounter{figure}{1}
    \begin{figure}
    \centering
    \includegraphics[width=8.2cm]{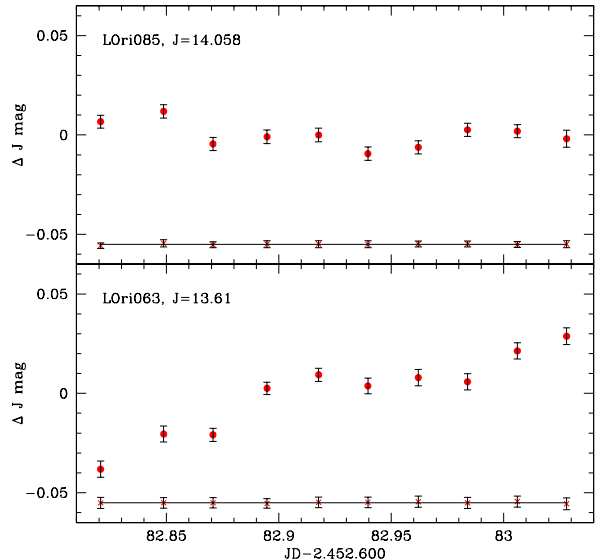}
 \caption{
Light-curves for two low mass stars belonging to the Lambda Orionis cluster.
These objects have been classified as Classical TTauri stars based on their location
on infrared  color-color diagrams, using Spitzer data. The data for the comparison stars
are shown at the bottom of each panel.
}
 \end{figure}

\subsection{Analysis of the variability}

The J-band light-curves of the bona-fide members are displayed in Figures 2--4.
The first diagram displays the curves for two low mass stars, whereas the second 
shows the data for two probable brown dwarfs. The last figure (Figure 4) 
corresponds to an object with a mass around the planetary mass limit 
($\sim$14 M$_{\rm Jup}$).
These light-curves have been constructed
subtracting the magnitudes of the targets from a non-variable, artificial
comparison star. 

 In order to construct the artificial comparison or reference star  we
have analyzed the light-curves of field stars with magnitudes as 
similar as possible  to those of the targets.
 From these objects, we have selected those that do not display
significant variability over the five hours of observations
The reference star is built by selecting field stars in a 
range of $\pm$2 mag around the problem star. After several 
iterations to remove variables in the field, we retain 
between 5 and 7  stars with no apparent variability. 
On average, the standard deviation of their brightness is below 0.008 
magnitudes.  We have weighted them according to their observational errors
and computed their average magnitude.
We have used a 3$\sigma$ as our variability criterion.
  In all figures, the light-curves of the comparison stars together with their mean
J-band magnitudes are shown below the
light-curves of the targets.


\subsection{Variability on the context of the evolutionary status}

Figure 5 displays the HR diagram for our targets.  Bolometric
luminosities were derived after the $I$ magnitudes the bolometric
corrections by Comer\'on et al. (2000), a distance modulus of 8.010 mag
and A$_I$=0.223.  Similar  values can be derived from $J$ and $K$ using
other bolometric corrections.  Effective temperatures come from Bessell
et al. (1991) and the $(R-I)$ color.  Note that the T$_{\rm eff}$ scale change
very much the Premain-sequence locus.  For instance, the scale by
Leggett (1992) would move the cluster sequence close to the 1 Myr
isochrone. Moreover, error-bars for a given  temperature scale
 can be close to 150-200 K.
  For the dataset represented in the diagram, the 5 Myr
Baraffe et al. (1998) fits the location within the errors.  The figure
also includes the D-instability strip (delimited by thick, dashed lines),
 which appears due to deuterium burning during
the first million years of pre(sub)stellar evolution.

LOri085 and LOri063 can be classified Class II objects based on
Spitzer/IRAC data (Barrado y Navascu\'es et al. 2006). 
For the first star, it seems that the variability, if present, is very small, at the
same level as in the case of the comparison star
(in our dataset, $\sigma$$_{\rm LOri085}$=0.007 mag, whereas $\sigma$$_{\rm comp}$=0.004 mag).
However, there is an obvious change in the case of LOri063.
Its variability can be due either to spots or changes in the
accretion rate. In the first cases the variability would be essentially periodic
(with a range from few to hundred of hours), whereas in the second case the 
light-curve would lack periodicity. Our dataset do not allow us to distinguish between
these possibilities.

LOri134 and LOri135 seem to be  bona-fide brown dwarfs 
 based on photometry and their spectral type. 
Based on Spitzer/IRAC data, they can be classified as Class II and Class III
objects, respectively, although LOri135 might have IR excess at 3.6 and 4.5 micron.
At the timescale of our photometric monitoring, the light-curves 
seem to be very different.

\setcounter{figure}{2}
    \begin{figure}
    \centering
    \includegraphics[width=8.2cm]{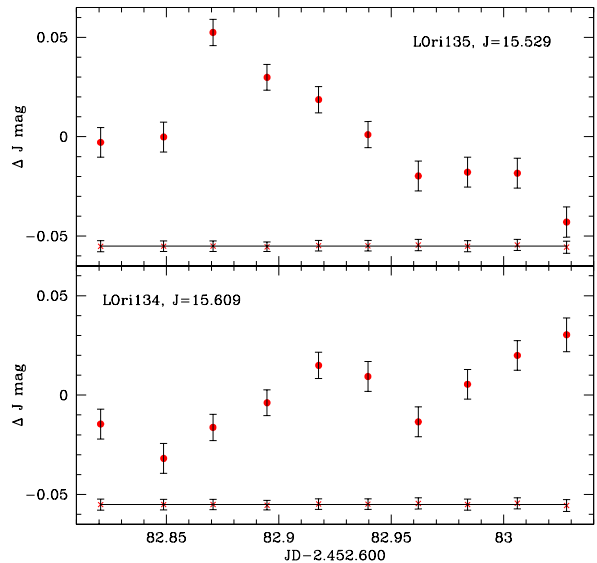}
 \caption{
Light-curves for two brown dwarf candidates.
}
 \end{figure}

 LOri167 is a cluster member based on its optical and near IR data and its
spectral energy distribution (from the $R$ band to 8.0 micron, 
Barrado y Navascu\'es et al. 2006). It is located close or within the
deuterium string instability in a HR diagram (Figure 5). 
The question is: Are we detecting pulsation? The light-curve suggests a 
period of about one  hour. However,  error-bars are large and the 
time coverage is not optimal to confirm this suggestive possibility. 
Note that the theoretical prediction by Palla \& Baraffe (2005) gives
a period  P(D-instability)$\sim$1.5 h.
Moreover, for an age of 5 Myr, the radius of LOri167 would be 0.225-0.143 R$_\odot$ for
a 0.02 M$_\odot$ brown dwarf (NextGen or COND models by the Lyon
group).  The escape velocity for this object would be 173 or 231 
Km/s for each model, after applying the formulae
V$_{\rm escape}$=617.88$\times$[M(M$_\odot$)/R(R$_\odot$)]$^{0.5}$.  Therefore, the minimum
rotational period would be P$_{\rm rot,min}$=2$\pi$R/V$_{\rm escape}$, which
translates into 1.79 h or 0.75 h for for NextGen and COND models,
respectively.  Since for the mass and T$_{\rm eff}$ of LOri167, the COND models are
better suited to describe its properties, and due to its location in
the HR diagram (very close or within the D-instability strip), it
seems that it is a real possibility that we have discovered a
D-burning pulsation. In any case, as stated before, this is just a possibility, 
and other explanations, more conventional, are  still possible (such as 
variability due to spots or  clouds in the atmosphere of the object).
Additional data, covering several nights and with better phtometric accuracy,
are needed.

\setcounter{figure}{3}
    \begin{figure}
    \centering
    \includegraphics[width=8.2cm]{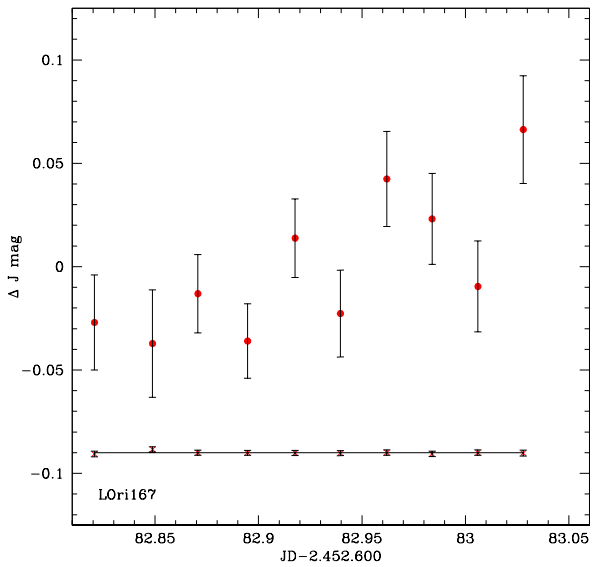}
 \caption{
Light-curve for LOri167. If member, this object would have a mass slightly above the  the
deuterium burning limit, located  at 0.013 M$_\odot$. Due to the cluster age, pulsations might be present.
}
 \end{figure}

\section{Conclusions}

By photometrically monitoring in the $J$ band a sample of low mass members of the
Lambda Orionis cluster, both stars and substellar objects, and using differential
photometry,  we have derived the light-curves in a five-hour time-span. 
We have detected variability in some of them, including a Classical TTauri star,  two
brown dwarfs and  a very faint object with 
J=17.8 mag, which corresponds to a mass close to the planetary mass domain 
for cluster members. This variability, with an amplitude of about 0.1 mag,
 might be due either to rotation, accretion or to instabilities --ie., pulsation-- 
due to the  deuterium burning, as proposed by Palla \& Baraffe (2005) for very low mass objects.

\setcounter{figure}{4}
    \begin{figure}
    \centering
    \includegraphics[width=8.2cm]{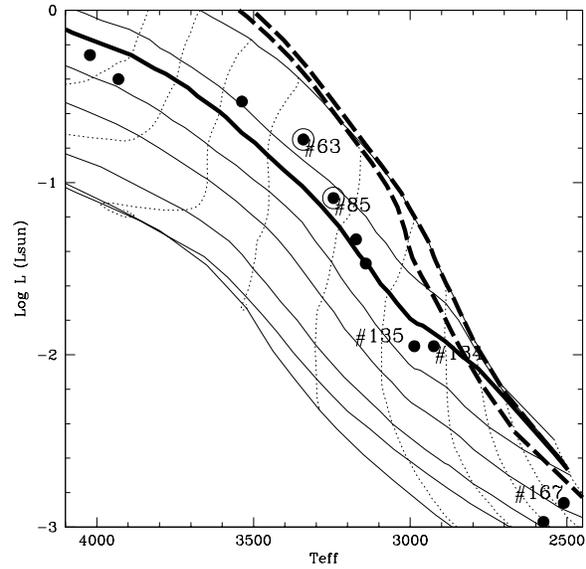}
 \caption{
HR diagram for our cluster candidates. We have included  isochrones 
(from top to bottom 1, 3, 5, 10, 20, 50, 100 and 1000 Myr, solid lines)
  and evolutionary tracks 
(from left to right  0.02, 0.03, 0.04, 0.06, 0.1, 0.2, 0.4, 0.6, 0.8, 1.0 M$_\odot$, dotted lines)
from Baraffe et al. (1998). The 5 Myr isochrone has been highlighted.
 The area delimited by the two thick dashed lines correspond to the deuterium instability, 
where pulsation might occur.
}
 \end{figure}

\acknowledgements
Based on data collected by the Willian Herschel Telescope at the
 Roque de los Muchachos observatory, La Palma, Spain.
We are indebted to F. Palla, I. Baraffe, and M. Fern\'andez,  and to the 
referee,  J. Caballero.

\end{document}